\documentclass[floatfix, amsmath,reprint, amssymb,aps,prb,twocolumn, superscriptaddress]{revtex4-2}
\usepackage{mathtools,amsmath,graphicx,units}
\usepackage{dcolumn}
\usepackage{bm}
\usepackage{xcolor}
\usepackage{physics}
\usepackage{float}
\usepackage{epstopdf}
\usepackage{enumitem}
\usepackage[normalem]{ulem}
\usepackage[plainpages=false,pdfpagelabels,colorlinks=true,linkcolor=red,urlcolor=blue,citecolor=blue,pdftitle={Title},pdfauthor={},pdfdisplaydoctitle=true,pdfduplex=DuplexFlipLongEdge]{hyperref}
\usepackage{cleveref}
\usepackage[T1]{fontenc}
\usepackage[makeroom]{cancel}

\begin{document}

\title{Observation of prethermalization in weakly nonintegrable unitary maps}
\author{Xiaodong Zhang}
\email{xiaodongzhang2021@gmail.com}
\affiliation{Center for Theoretical Physics of Complex Systems, Institute for Basic Science, Daejeon 34126, Republic of Korea}
\affiliation{
Lanzhou Center for Theoretical Physics, Key Laboratory of Theoretical Physics of Gansu Province, \\
and Key Laboratory of Quantum Theory and Applications of MoE, Lanzhou University, Lanzhou, Gansu 730000, China
}
\author{Gabriel M.~Lando}
\email{gmlando@ibs.re.kr}
\affiliation{Center for Theoretical Physics of Complex Systems, Institute for Basic Science, Daejeon 34126, Republic of Korea}
\author{Barbara Dietz}
\email{bdietzp@gmail.com}
\affiliation{Center for Theoretical Physics of Complex Systems, Institute for Basic Science, Daejeon 34126, Republic of Korea}
\affiliation{Basic Science Program, Korea University of Science and Technology (UST), Daejeon 34113, Republic of Korea}
\author{Sergej Flach}
\email{sergejflach@googlemail.com}
\affiliation{Center for Theoretical Physics of Complex Systems, Institute for Basic Science, Daejeon 34126, Republic of Korea}
\affiliation{Basic Science Program, Korea University of Science and Technology (UST), Daejeon 34113, Republic of Korea}

\date{\today}
    
\begin{abstract}
We investigate prethermalization by studying the statistical properties of the time-dependent largest Lyapunov exponent $\Lambda(t)$ for unitary-circuit maps upon approaching integrability. We follow the evolution of trajectories for different initial conditions and compute the mean $\mu(t)$ and standard deviation $\sigma(t)$ of $\Lambda(t)$.
Thermalization implies a temporal decay $\sigma \sim t^{-1/2}$ at a converged finite value of $\mu$. We report prethermalization plateaus that persist for long times where both $\mu$ and $\sigma$ appear to have converged to finite values, seemingly implying differing saturated Lyapunov exponent values for different trajectories. The lifetime of such plateaus furnishes a novel time scale characterizing the thermalization dynamics of many-body systems close to integrability. We also find that the plateaus converge to their respective thermal values for long enough times. 
\end{abstract}
\maketitle

\noindent Dedicated to Professor Alexander Kovalev from B.~Verkin Institute for Low Temperature Physics and Engineering of the NASU (Kharkiv, Ukraine) on the occasion of his 80th birthday.

\section{Introduction}

The study of thermalization usually starts from the Gibbs assumption of equal probability for each microstate, and assumes that the system displays ergodicity, i.e. infinite time averages have to be equal to phase space (ensemble) averages \cite{gibbs1906scientific}. 
A standard approach to investigate thermalization dynamics then consists of choosing observables and extracting ergodization time scales on which their time averages converge to their ensemble averages. 
Since these observables are functions of phase-space coordinates, their phase-space averages are not correlated with the ergodization time, which can diverge when tuning the system parameters towards an integrable limit.
Thus, the ambiguity in the choice of observables ends up 
in a multitude of different ergodization times. 

To bypass such ambiguity, an alternative approach consists in computing Lyapunov exponents - either the largest or the entire spectrum \cite{skokos_lyapunov_2010}. The inverse of a Lyapunov exponent provides a unique time scale which characterizes the exponential decay of correlations in the system. The computation reduces again to the time averaging of a certain quantity along a trajectory, which in this case can no longer be interpreted as an observable. Indeed, it is uniquely defined from the system's Hamiltonian and keeps track of correlations along the trajectory. A remarkable property of these quantities is that the resulting Lyapunov exponents are differential invariants \cite{eichhorn2001transformation}, such that there is no ambiguity in their associated time scales, e.g. their values being inversely proportional to the ergodization times of their corresponding Lyapunov observables.

The measurement of Lyapunov times and ergodization times of properly chosen observables
provide new insights into the slowing down of thermalization upon approaching integrable limits \cite{danieli_dynamical_2019,mithun_dynamical_2019,mithun_fragile_2021,lando2023thermalization,malishava_lyapunov_2022,malishava_thermalization_2022,
zhang2024thermalization}. The Lyapunov spectrum scaling shows universality and establishes different classes of weakly nonintegrable perturbations. These perturbations connect the actions, which are conserved for the integrable limit model, into long range networks (LRN) or short range networks (SRN). The different network classes demonstrate unique correlations between the scalings of the Lyapunov spectrum and the ergodization time of the above actions \cite{malishava_lyapunov_2022,malishava_thermalization_2022}. The LRN scaling is roughly characterized by one diverging time scale (e.g. the smallest Lyapunov time obtained by inverting the largest Lyapunov exponent), i.e. all other thermalization time scales diverge proportionally. The SRN scaling is characterized by a second diverging quantity - the exponent which controls the decay of the Lyapunov spectrum. This exponent and its divergence are believed to be connected to the divergence of the distance between chaotic multiplets of actions \cite{danieli_dynamical_2019,mithun_dynamical_2019,mithun_fragile_2021}.
The distance here is measured in action space using the network metrics, and corresponds to a distance in a real physical space for most models.

The proximity to an integrable limit results in slow thermalization, which is particularly prone to intermediate {\sl prethermalization} dynamics. As a key phenomenon in nonequilibrium dynamics, prethermalization has attracted significant attention across various research fields \cite{PhysRevB.84.054304,science.1224953,PhysRevLett.115.180601,PhysRevB.95.024302,sciadv.1700672,Le2023,PhysRevD.107.086001,Birnkammer2022,PhysRevB.109.195140,PhysRevB.109.054202,PhysRevB.98.104303,PhysRevX.10.021046,PhysRevLett.122.080603}. 
Many studies use the existence of additional integrals of motion (e.g. spin systems, many-body quantum systems with conserved particle numbers or their classical analogues if available) or at least the existence of almost conserved quantities (e.g. the energy in certain realizations of Floquet systems) to enforce slow relaxation. Typically a quench is performed, and a particular observable is picked, to follow the fluctuations of the latter and to extract either slow or even the absence of relaxation over significant times. A few studies pick random initial states, and again choose a particular observable to be followed. Others even take integrable systems and use generalized Gibbs-ensemble concepts. Most of the investigations are performed in the quantum regime and few of them in the classical one. 
The previously discussed ambiguity in the choice of observables appears to be a limiting factor in characterizing thermalization and therefore also prethermalization, attesting to how tricky such characterization can be. Thus, in view of their invariance, it is interesting to search for prethermalization characteristics of typical initial states through the measurement of Lyapunov exponents. 

In the prethermal regime, observables behave as if converged for temporal windows of varying length before finally relaxing to their true asymptotic values. One way to observe such phenomenon is to start from a non-typical initial state for which the dynamics is trapped in a near-regular part of phase space for long times. A prominent example is the Fermi-Pasta-Ulam-Tsingou paradox (FPUT) \cite{FPUTpreprint,FPUT-CP}. Exciting a single mode (action) of a weakly nonintegrable anharmonic chain results in surprisingly long times to reach equipartition, with Lyapunov times being much shorter, suggesting some converged Lyapunov spectrum for a system still being far from convergence \cite{1997PhRvE..55.6566C}. Another instance of prethermal dynamics that does not require the use of special initial states is the previously discussed SRN regime of a weakly nonintegrable system, which is the approach pursued here. 

We study thermalization dynamics in a one-dimensional system of nonintegrable nonlinear unitary circuit maps \cite{malishava_lyapunov_2022,malishava_thermalization_2022}. These maps conserve the total norm, similar to the conservation of total energy for continuous-time Hamiltonian dynamics. Note that technically unitary dynamics can be mapped back onto a stroboscobic Poincar\'e-like map of some underlying Hamiltonian system, which conserves both norm and energy. What matters is the bulk of evidence that the thermalization dynamics of unitary maps considered here show full similarity to the conventional thermalization dynamics of Hamiltonian systems \cite{constantoudis_nonlinear_1997,mithun_weakly_2018,danieli_dynamical_2019,mithun_dynamical_2019,mithun_fragile_2021,lando2023thermalization,zhang2024thermalization}.
We quantify the temporal convergence properties of Lyapunov exponents using trajectories with randomly chosen initial values. We observe the tantalizing convergence of Lyapunov exponents along such randomly chosen trajectories for time windows of varying duration, with the theoretically expected convergence to a unique and trajectory-independent value taking place only at much longer times.

\section{Model}

We employ the nonlinear unitary circuit map used in Refs.\cite{malishava_lyapunov_2022,malishava_thermalization_2022,zhang2024thermalization}. It is a one-dimensional lattice consisting of \( N \) unit cells, where each unit cell is labeled by an odd site \( n \) and its corresponding even site \( n+1 \). Here, \( n \) takes odd values (\( n = 1, 3, 5, \dots, 2N-1 \)), and \( n+1 \) is the next even site (\( n+1 = 2, 4, 6, \dots, 2N \)). Each site \( n \) is represented by a complex component \( \psi_n \), and the initial state of the lattice is described by the vector \(\boldsymbol{\psi}=\left(\psi_1, \psi_2, \psi_3, \psi_4, \dots, \psi_{2N-1}, \psi_{2N}\right)\), representing all the sites in the lattice. The system evolves in a phase space of dimension \( 4N \), as each complex component is one degree of freedom which contributes two real variables. The evolution follows a deterministic trajectory defined by the initial conditions and is achieved through iterative applications of the unitary map,
\begin{equation}
\hat{U}_n = \hat{G}_n \hat{C}_{n-1,n} \hat{C}_{n,n+1},
\label{eq1}
\end{equation}
with each iteration corresponding to one time step.
\begin{figure}
\includegraphics[width=0.45\textwidth]{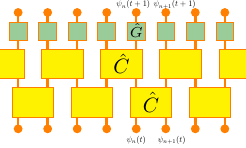}
    \caption{A schematic representation of the unitary circuits map. It evolves from bottom to top and consists of alternating large yellow blocks and small light green blocks, representing unitary matrices \(\hat{C}\) parameterized by the angle \(\theta\) and local nonlinearity generating maps \(\hat{G}\) parameterized by the nonlinearity strength \(g\), respectively. The states \(\psi_{n}(t)\) and \(\psi_{n+1}(t)\) evolve to \(\psi_{n}(t+1)\) and \(\psi_{n+1}(t+1)\), respectively, through subsequent applications of these transformations. 
    One time step contains three substeps. 
    }
    \label{map}
\end{figure}
The unitary map \( \hat{U}_n \) is constructed from two key components, namely two successive rotation operations and a nonlinear operation as illustrated in Fig. \ref{map}. The rotation operations, \( \hat{C} \), are unitary transformations that act on pairs of neighboring sites \( n \) and \( n+1 \). Specifically, they are defined as,
\begin{equation}
      \hat{C}_{n,n+1}\begin{pmatrix}\psi_n(t)\\\psi_{n+1}(t)\end{pmatrix}=\begin{pmatrix}\cos\theta&\sin\theta\\-\sin\theta&\cos\theta\end{pmatrix}\binom{\psi_n(t)}{\psi_{n+1}(t)}.
\label{eq2}
\end{equation}
In addition to the rotation operations, the map includes a nonlinear operation \( \hat{G}_n\), which introduces site-dependent phase shifts that are proportional to the norm \(|\psi_{\nu}(t)|^2\) on site $\nu,\, \nu=n,n+1 $. The nonlinear operation is defined as,
\begin{equation}   \hat{G}_n\psi_\nu(t)=e^{ig|\psi_{\nu}|^2}\psi_{\nu}(t).
\label{eq3}
\end{equation}
By combining the two rotation operations \( \hat{C}_{n-1,n} \) and \( \hat{C}_{n,n+1} \) with the nonlinear operation \( \hat{G}_{n} \), we derive the evolution equations,
\onecolumngrid 
\vspace{0.5cm}
\noindent\rule{0.5\textwidth}{0.4pt}
\hspace{-7pt}
\raisebox{0cm}[0pt][0pt]{\rule{0.4pt}{0.4cm}}
\begin{equation}
\begin{aligned}
\psi_{n}(t+1) &= 
 e^{i g |\varphi_{n}(t)|^2 }\varphi_{n}(t),\quad \varphi_{n}(t)=\left( \sin^2 \theta \psi_{n-2}(t) - \cos \theta \sin \theta \psi_{n-1}(t) + \cos^2 \theta \psi_n(t) \right.\left. + \sin \theta \cos \theta \psi_{n+1}(t) \right)\\
\psi_{n+1}(t+1) &= 
e^{i g |\varphi_{n+1}(t)|^2 }\varphi_{n+1}(t),\quad \varphi_{n+1}(t)=\left( -\sin \theta \cos \theta \psi_n(t) + \cos^2 \theta \psi_{n+1}(t) \right. \left. + \sin \theta \cos \theta \psi_{n+2}(t) + \sin^2 \theta \psi_{n+3}(t) \right),
\label{eq1.4}
\end{aligned}
\end{equation}
\noindent\hspace*{\columnwidth}\hspace*{-0.5\textwidth}\rule{0.5\textwidth}{0.4pt}
\hspace*{0.5\textwidth}\raisebox{0cm}[0pt][0pt]{\rule{0.4pt}{0.4cm}}
\vspace{0.5cm}
\twocolumngrid
\noindent where \( \varphi_{n} \) and \( \varphi_{n+1} \) denote the states 
\( \psi_{n} \) and \( \psi_{n+1} \), respectively, after applying two rotation operations \(\hat{C}\).
In the simulations, we use periodic boundary conditions $\psi_{2N+1}=\psi_{1}$. 
The map ensures that the total squared norm \( \mathcal{A} = \sum_m |\psi_m(t)|^2 \) is a conserved quantity. The local norm density \( |\psi_m|^2 \) is drawn from a Gibbs distribution $\rho (x) = 2{\rm e}^{-2x}$ with an average squared-norm density of \( a=\mathcal{A}/(2N)=\frac{1}{2} \), and the phases are chosen from a distribution which is uniform on the interval \([0, 2\pi]\). 
This map possesses two distinct integrable limits $g=0$ (linear case) and $\theta=0$ (decoupled sites). The limit $\theta=0$ corresponds to the case where the sites are decoupled, and the norm of each site is conserved, resulting in \(2N\) independent conserved quantities. The squared norms here can be regarded as the analog of the actions in time-continuous Hamiltonian systems. In the limit $g=0$ the nonlinearity vanishes, but the coupling between the sites remains, and the norms of the corresponding normal modes are conserved. Weak perturbations off the integrable limits result in LRN ($g \ll a,\theta$) and SRN ($\theta \ll a,g $) universality classes \cite{malishava_lyapunov_2022,malishava_thermalization_2022}
(see also Appendix \ref{Resonance}). 

\section{Lyapunov exponent computation}

We determine the time evolution of deviation vectors to obtain the time-dependent largest Lyapunov exponent. To derive the equations of motion for the deviation vectors, we decompose the trajectory $\boldsymbol{\Psi}(t)$ into an unperturbed trajectory $\boldsymbol{\psi}(t)$ and a deviation $\boldsymbol{W}(t)$. Thus, the trajectory can be expressed as $\boldsymbol{\Psi}(t) = \boldsymbol{\psi}(t) + \boldsymbol{W}(t)$. For convenience, we define the two operators, \(\alpha_n\) and \(\alpha_{n+1}\),
\onecolumngrid 
\vspace{0.5cm}
\noindent\rule{0.5\textwidth}{0.4pt}
\hspace{-7pt}
\raisebox{0cm}[0pt][0pt]{\rule{0.4pt}{0.4cm}}
\begin{equation}
\begin{aligned}
\alpha_n X_n &\coloneqq \sin^2 \theta X_{n-2} - \cos \theta \sin \theta X_{n-1} + \cos^2 \theta X_n+ \sin \theta \cos \theta X_{n+1}, \\
\alpha_{n+1} X_{n+1} &\coloneqq -\sin \theta \cos \theta X_n + \cos^2 \theta X_{n+1} + \sin \theta \cos \theta X_{n+2} + \sin^2 \theta X_{n+3}.
\label{operator}
\end{aligned}
\end{equation}
\noindent\hspace*{\columnwidth}\hspace*{-0.5\textwidth}\rule{0.5\textwidth}{0.4pt}
\hspace*{0.5\textwidth}\raisebox{0cm}[0pt][0pt]{\rule{0.4pt}{0.4cm}}
\vspace{0.5cm}
\twocolumngrid

Expanding the nonlinear exponential term in the equations \sout{}for $\Psi_{n}(t)$ and $\Psi_{n+1}(t)$ followed by a linearization in $W_\nu(t)$ with $\nu=n,n+1$, we derive its linearized evolution equation \\
\begin{align}
W_\nu(t+1) &= e^{ig|\alpha_{\nu} \psi_{\nu}(t)|^2} \cdot (\alpha_\nu W_{\nu}(t) + i g \Delta_\nu(t) \cdot \alpha_\nu \psi_{\nu}(t)), \notag \\
\Delta_\nu(t) &= \alpha_\nu \psi_{\nu}(t) \alpha_\nu W_{\nu}(t)^{*} + \alpha_\nu \psi(t)_{\nu}^{*} \alpha_\nu W_{\nu}(t). 
\label{eq3.7}
\end{align}
We are only interested in the largest time-dependent Lyapunov exponent $\Lambda(t)$, which is obtained from the running time-average of the Lyapunov observable $r(t) = \ln\|\boldsymbol{W}(t)\|$, where \(\|\boldsymbol{W}(t)\|\) is the length of the deviation vector $\boldsymbol{W}(t)$. Accordingly, we compute at each time step the Lyapunov observable $r(t)$ and then
the temporal average $\Lambda(t)= \frac{1}{t} \sum_{\tau=0}^{t} r(\tau)$. After each calculation step, $\boldsymbol{W}(t)$ needs to be normalized. 
It follows that $\Lambda(t)$ is a running time average of the Lyapunov observable $r(t)$. In contrast to usual observables, $r(t)$ can not be simply expressed through a function of the phase space coordinates. It keeps correlations and memory along the trajectory as the trajectory evolves. Attempts to replace its time average by a procedure of phase space averaging are notoriously complicated, and usually only possible in certain limiting regimes \cite{1995PhRvL..74..375C,casetti_riemannian_1996}.
At the same time, and in the spirit of the ergodic theorem, the computational time average of $r(t)$ is usually saturating at its ergodization time and becomes time independent. 

Since we are only calculating the largest Lyapunov exponent, \( \boldsymbol{W}(t) \) is merely a random complex vector. The largest Lyapunov exponent represents the strongest exponential divergence in the system. Calculating only the largest exponent not only efficiently captures the key dynamical features but also reduces computational complexity, allowing simulations to extend to longer time scales.

\section{Statistical properties of Lyapunov observables}

To understand the statistical behavior of Lyapunov observables in systems approaching an integrable limit, we analyze their probability-density function (PDF) $P(r)$ of the Lyapunov observables $r(t)$ and associated statistical properties, such as the mean and standard deviation. We performed \(10^9\) iterations of the unitary map Eq.~(\ref{eq1}) yielding $10^9$ values of \(r(t)\) that were accounted for in the evaluation of their PDF.  Two distinct regimes are considered, namely LRN and SRN.
\begin{figure}[!htbp]
\includegraphics[width=0.45\textwidth]{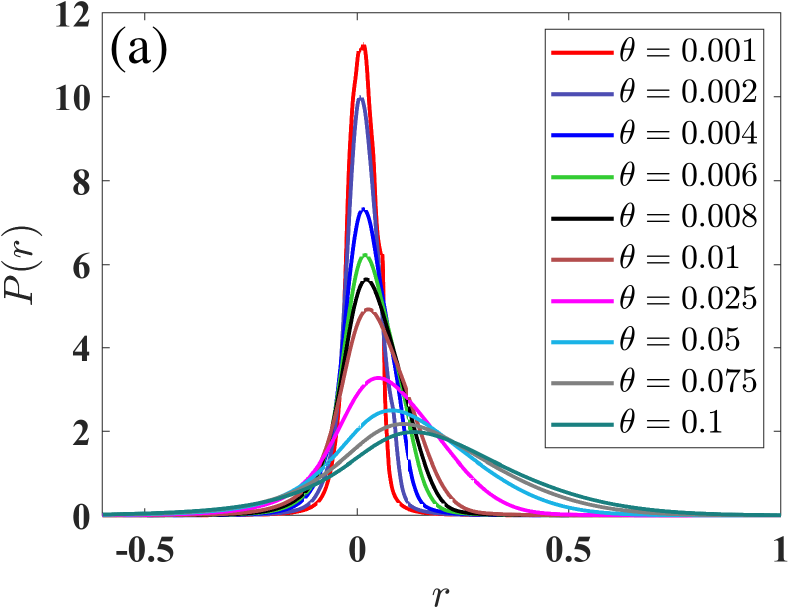}
\includegraphics[width=0.45\textwidth]{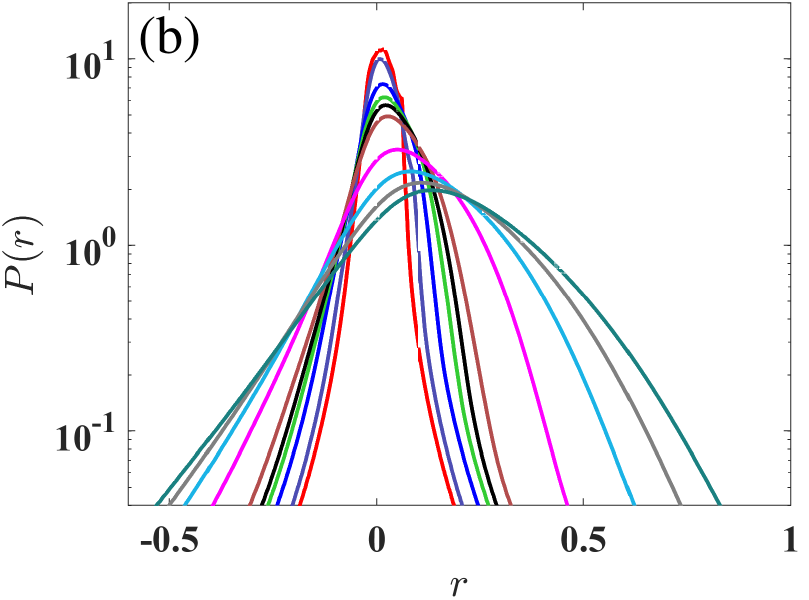}
    \caption{Probability-density function $P(r)$ of the Lyapunov observable \( r \) for the SRN case with \( g = 1 \). Panel (a) shows the PDF on linear scales, while panel (b) presents the same data with the \( y \)-axis plotted on a log\(_{10}\) scale. The colors corresponding to different values of \( \theta \) in (a) and (b) are shown in the legend of panel (a). The unitary map Eq.~(\ref{eq1}) is iterated \( 10^9 \) times, yielding \( 10^9 \) values for \(r(t)\) that were all used to calculate $P(r)$. The number of unit cells in the system is \( N = 100 \).}
    \label{PDFSRN}
\end{figure}

\begin{figure}[!htbp]
\includegraphics[width=0.46\textwidth]{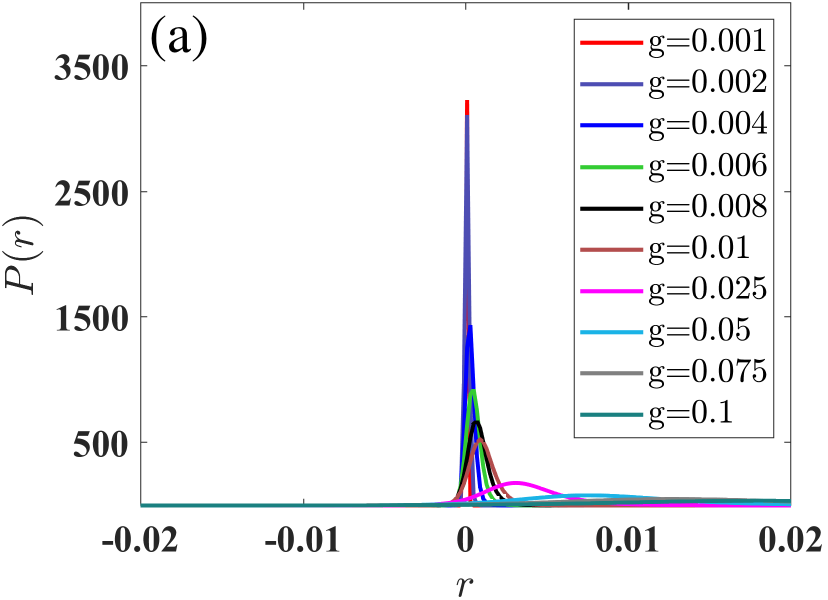}
\includegraphics[width=0.45\textwidth]{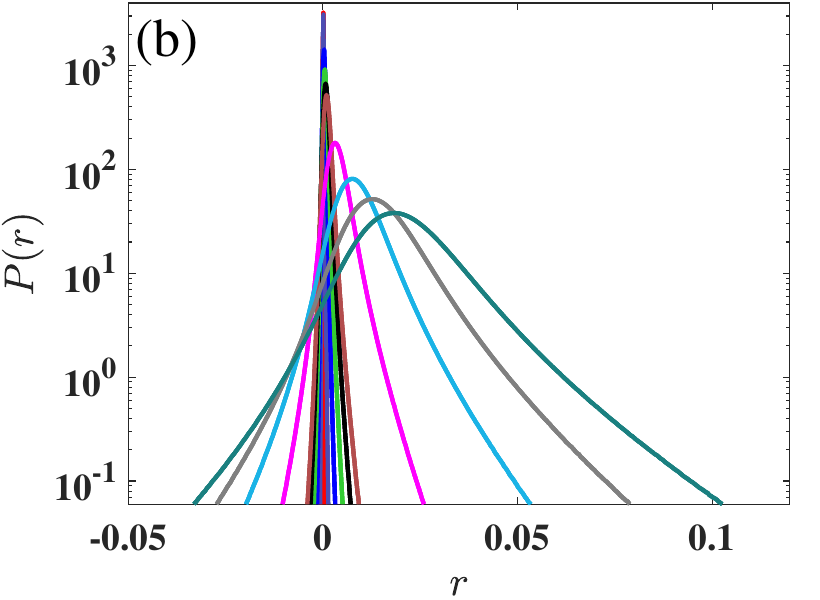}
    \caption{Same as Fig.~\ref{PDFSRN} for the LRN case with \( \theta = 0.33\pi \).  The colors corresponding to the different values of \( g \) in (a) and (b) are shown in the legend of panel (a).}
    \label{PDFLRN}
\end{figure}

Figure~\ref{PDFSRN} shows the PDF \( P(r) \) for the SRN case with \( g = 1 \) and various values of \( \theta \) ranging from \( 0.001 \) to \( 0.1 \). Panel (a) presents the PDF on a linear scale, while panel (b) uses a logarithmic scale (base \( 10 \)) on the \( y \)-axis. The results reveal a clear trend, namely as \( \theta \) decreases, \( P(r) \) becomes increasingly symmetrically concentrated around \( r = 0 \), with the peak height near \( r = 0 \) growing significantly. This behavior is particularly evident in panel (b), where the logarithmic scale highlights the increase of the peak height as \( \theta \) approaches smaller values. Physically, this reflects the system approach toward the marginal stability regime, where the Lyapunov observable \( r(t) \) remains close to zero for longer durations. At larger values of \( \theta \), the distribution broadens, indicating stronger deviations from zero and a higher degree of dynamical instability. The slow yet observable steepening of peak of \( P(r) \) near \( r = 0 \) for small \( \theta \) highlights the system’s tendency to exhibit weaker chaotic dynamics as it approaches the integrable limit.

Figure~\ref{PDFLRN} shows the PDF \( P(r) \) for the LRN case with \( \theta = 0.33\pi \). Panel (a) presents the PDF on a linear scale, while panel (b) shows the same data with the \( y \)-axis plotted on a logarithmic scale (base \( 10 \)). The behavior of \( P(r) \) for different values of \( g \) (ranging from \( 0.001 \) to \( 0.1 \)) is qualitatively consistent with the SRN case shown in Fig. 2. As \( g \) decreases, the distribution becomes increasingly peaked around \( r = 0 \), reflecting the same trend observed in SRN, namely the Lyapunov observable spends more time near zero when approaching smaller \( g \). 
Note, however, that the LRN PDF appears to become narrower much faster upon approaching its integrable limit, than the SRN PDF. To quantify these observations, we compute the mean and standard deviations of the PDFs.
\begin{figure}[!htbp]
\includegraphics[width=0.45\textwidth]{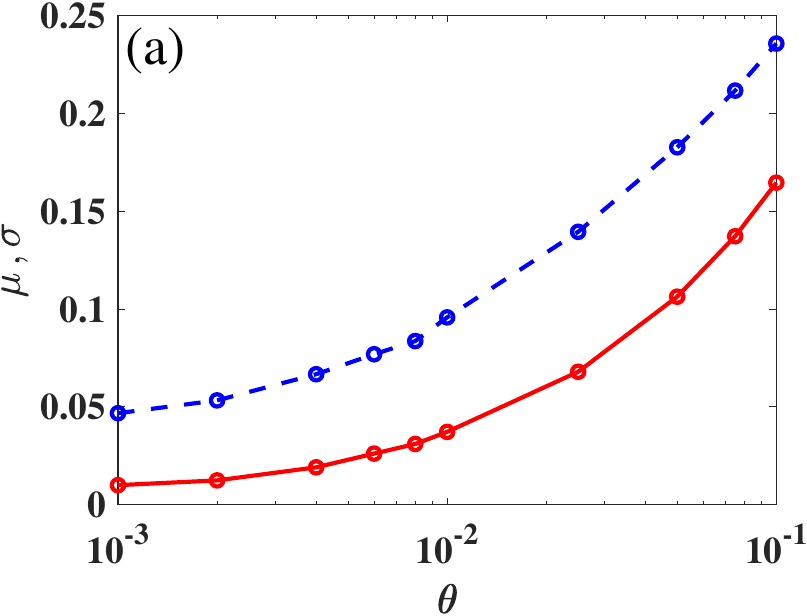}
\includegraphics[width=0.45\textwidth]{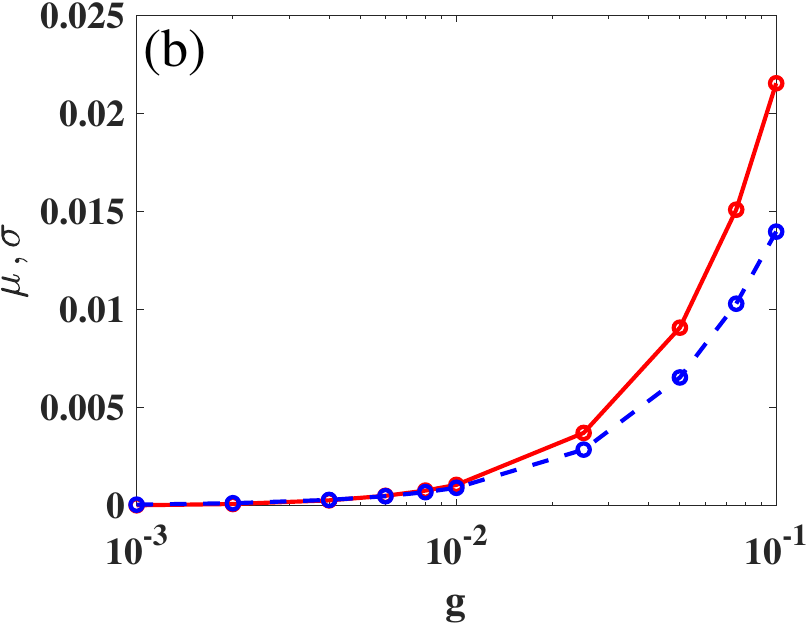}
    \caption{Average value, \( \mu \), and standard deviation, \( \sigma \), of the Lyapunov observable for the SRN and LRN cases, with both panels using a log\(_{10}\) scale on the \( x \)-axis. Panel (a) represents the SRN case with \( g = 1 \) and varying \( \theta \), while panel (b) shows the LRN case with \( \theta = 0.33\pi \) and varying \( g \). The red solid line represents the mean value (\( \mu \)), and the blue dashed line represents the standard deviation (\( \sigma \)). The circles correspond to the specific values of \( g \) or \( \theta \) used in the calculation. The unitary map Eq.~(\ref{eq1}) is iterated \( 10^9 \) times, yielding \( 10^9 \) values for \(r(t)\) that were all used to compute the mean and standard deviations. The system consists of \( N = 100 \) unit cells.}
    \label{muLO}
\end{figure}
Figure \ref{muLO} shows the average value and standard deviation of the Lyapunov observable \( r(t) \), with panel (a) exhibiting the SRN case and panel (b) showing the LRN case. In the SRN case, as \( \theta \) approaches smaller values, both \( \mu \) and \( \sigma \) decrease. However, the standard deviation remains larger than the mean while both \( \mu \) and \( \sigma \) appear to saturate at a nonzero value. Extensive further computations (see Appendix \ref{LYAPOBS}) 
indicate that both mean and standard deviation eventually tend towards zero upon approaching the integrable limit, yet this process appears to be rather slow. This behavior indicates that as \( \theta \to 0 \), the Lyapunov observable shows anomalous nonzero fluctuations, a possible prerequisite to prethermalization. 

For the LRN case, as \( g \to 0 \), both \( \mu \) and \( \sigma \) tend to zero much faster than in the SRN case.  The standard deviation quickly approaches the mean and both quantities decrease rapidly towards zero.  Note that the standard deviation reaches values by a factor of $10^{-3}$ smaller than for the SRN case, and the mean reaches values by a factor $10^{-2}$ smaller than for the SRN case. As we further reduce $g$ we start to see anomalies similar to those for the SRN case (see appendix \ref{LYAPOBS}).

We conclude this section with the observation that the statistics of Lyapunov observables in the LRN regime appears to show expected features - no anomalous fluctuations, and fast diminishing distribution width and peak position upon approaching the integrable limit. On the contrary, in the SRN regime we observe anomalous fluctuations, and mean and standard deviation seemingly frozen at nonzero values which are 100-1000 times larger than the corresponding numbers from the LRN regime. 

\section{Observation of Prethermalization}

The statistical analysis of the Lyapunov observable does not account for temporal correlations between the measured observables generated by the flow along a given trajectory. Prethermalization is expected to result in long-time periods with seemingly converged Lyapunov exponents whose values nevertheless depend on the initial conditions. That implies that temporal correlations do not decay up to times beyond such time windows. In this section we analyze the time evolution of \( \Lambda(t) \) for the SRN and LRN systems, i.e. the running time averages of the Lyapunov observable along a given trajectory.

\begin{figure}[!htbp]
\includegraphics[width=0.48\textwidth]{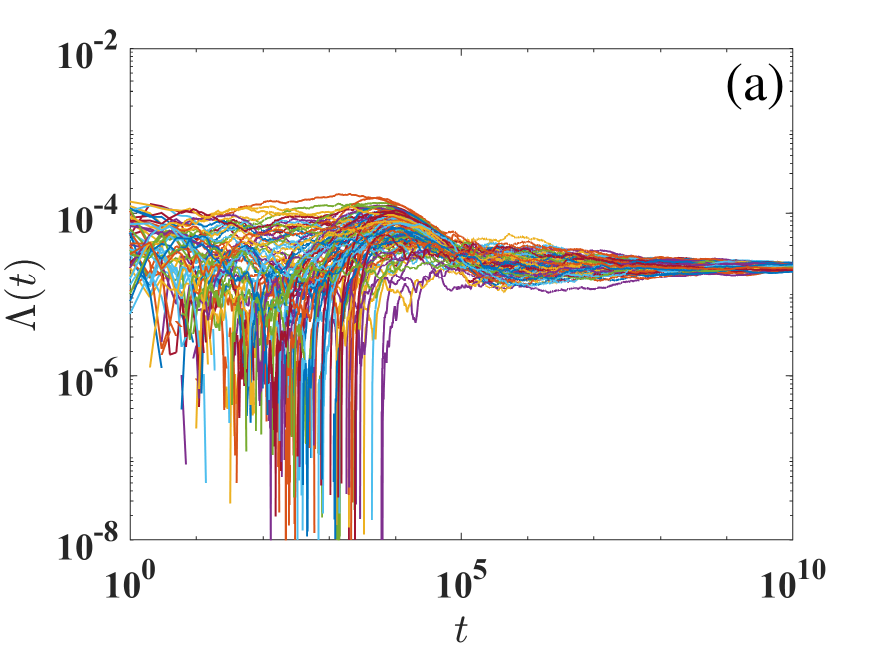}
\includegraphics[width=0.48\textwidth]{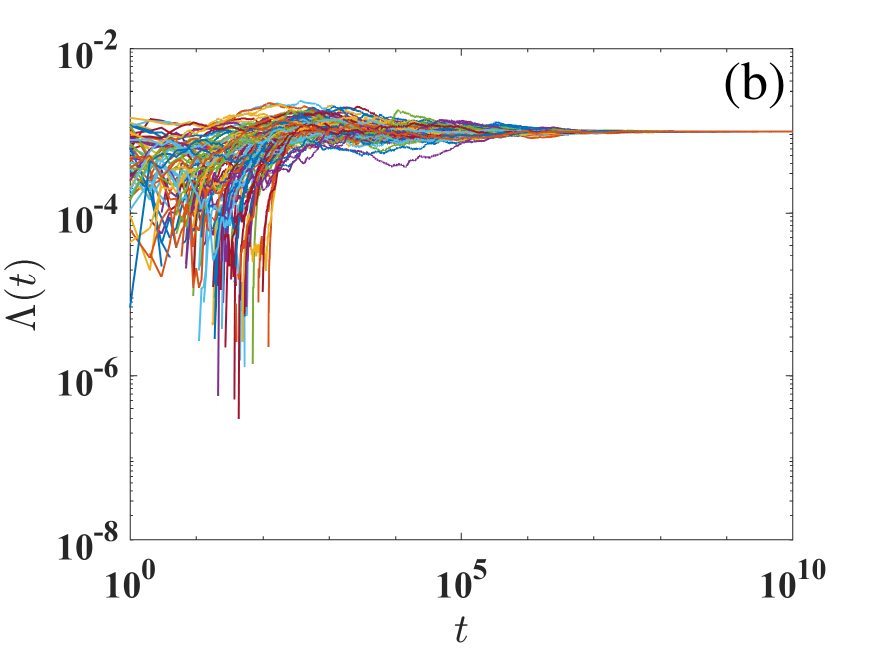}
    \caption{Time evolution of the largest Lyapunov exponent \(\Lambda(t)\) on a log\(_{10}\) scale for up to \(10^{10}\) iterations for  \(g=0.001\) (a) and \(g=0.01\) (b) in the LRN case for \(\theta = 0.33 \pi\). Each panel shows 100 trajectories obtained from randomly chosen  initial conditions. The system consists of \(N = 50\) unit cells.}
    \label{LRN_LLE_N50}
\end{figure}

\begin{figure}[!htbp]
\includegraphics[width=0.48\textwidth]{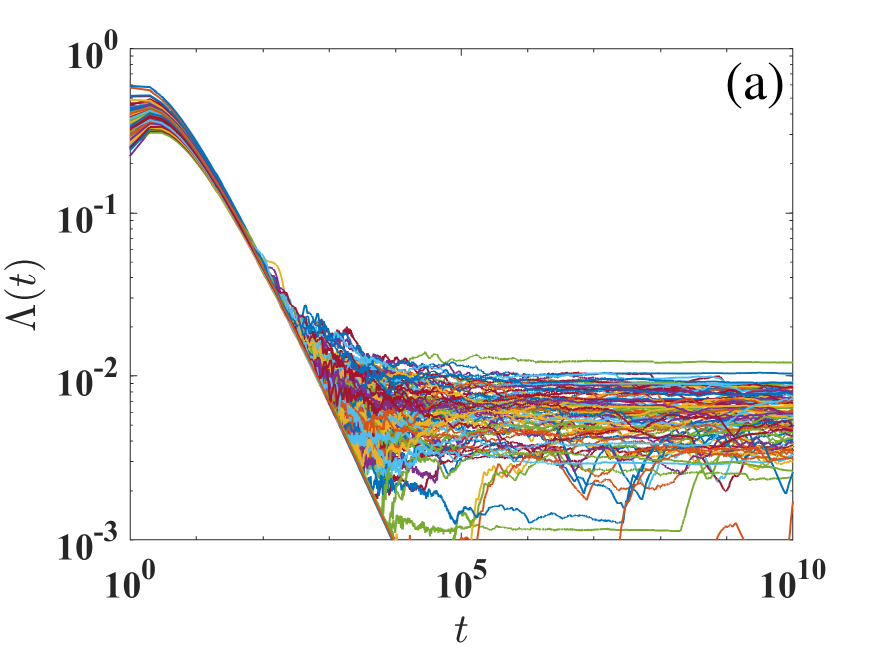}
\includegraphics[width=0.48\textwidth]{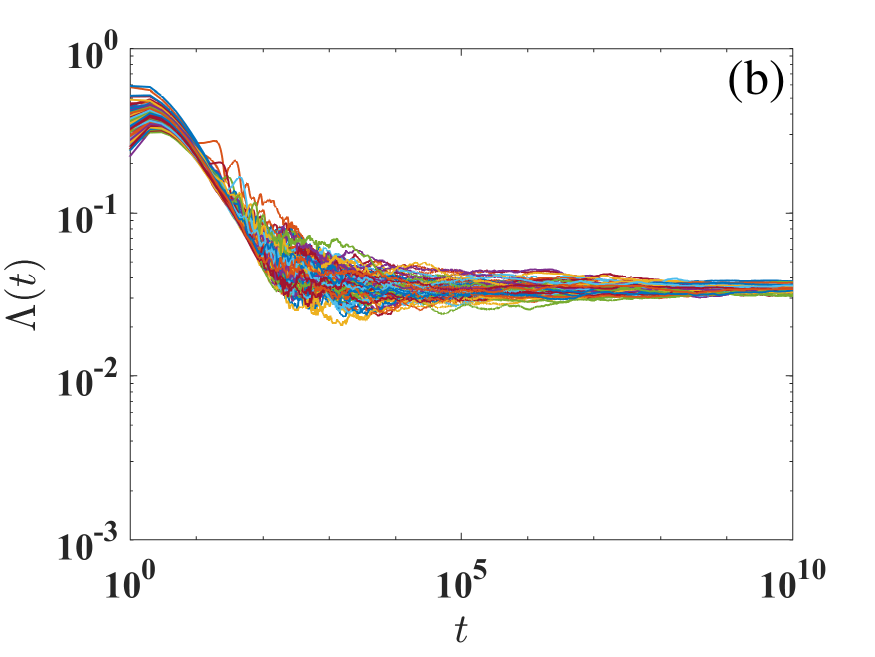}
    \caption{Same as Fig.~\ref{LRN_LLE_N50} for the SRN case with \(g = 1\) and $\theta=0.001$ (a)  and $\theta=0.01$ (b).}
    \label{SRN_LLE_N50}
\end{figure}
Figure~\ref{LRN_LLE_N50} shows the time evolution of the largest Lyapunov exponent \( \Lambda(t) \) for the LRN system with a fixed coupling parameter \( \theta = 0.33\pi \) for (\( g = 0.001, 0.01 \)), each for 100 different trajectories with random initial conditions. For $g=0.01$ the running time average quickly converges to a number which appears to be rather trajectory independent. This is the usual perception of a well thermalized ergodic system, with time averages of observables being independent of the chosen trajectory. Reducing the nonlinearity parameter to $g=0.001$ does not change the outcome, except that it delays the final curve saturation time and the saturation level, as discussed above. We conclude that for these parameter values, the LRN regime shows reasonable ergodic thermalization with an accompanying slowing down upon approaching the integrable limit. More data for intermediate values of $g$ are shown in Appendix \ref{LLEDIC}.

The SRN regime shows a qualitatively different result in Fig.~\ref{SRN_LLE_N50}. We fix the nonlinearity parameter at $g=1$ and show the time evolution of the largest Lyapunov exponent \( \Lambda(t) \) for the SRN system  for (\( \theta = 0.001, 0.01 \)), each for 100 different trajectories with random initial conditions. 
For $\theta=0.01$ the outcome is qualitatively similar to the LRN plots in Fig.~\ref{LRN_LLE_N50} (a), with reasonable thermalization and ergodicity. However, the plot for $\theta=0.001$ exhibits an extraordinary behavior. Different trajectories for different initial conditions appear to saturate, but at different values of the Lyapunov exponent. This clearly indicates that the system is not yet thermalized and ergodic within these time scales. We observe pre-thermalization, i.e. the system shows practically saturated Lyapunov exponents and finite-strength chaos, but the quantitative characteristics depend on the initial conditions. The only possibility is then that the seemingly saturated curves will ultimately merge into one proper asymptotic horizontal line for much larger times. More data for intermediate values of $\theta$ are shown in Appendix \ref{LLEDIC} and confirm this expectation.

\begin{figure}[!htbp]
\includegraphics[width=0.45\textwidth]{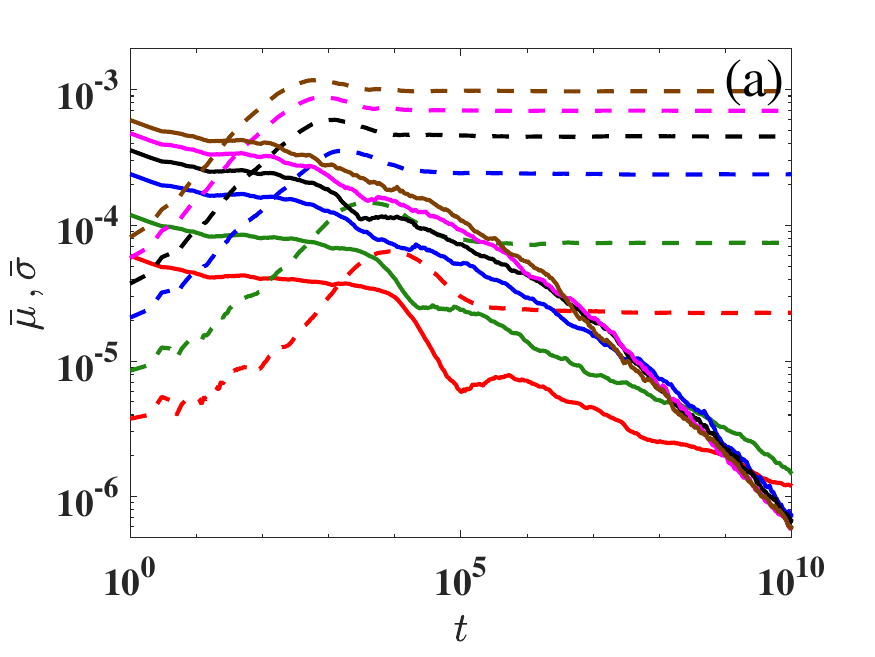}
\includegraphics[width=0.45\textwidth]{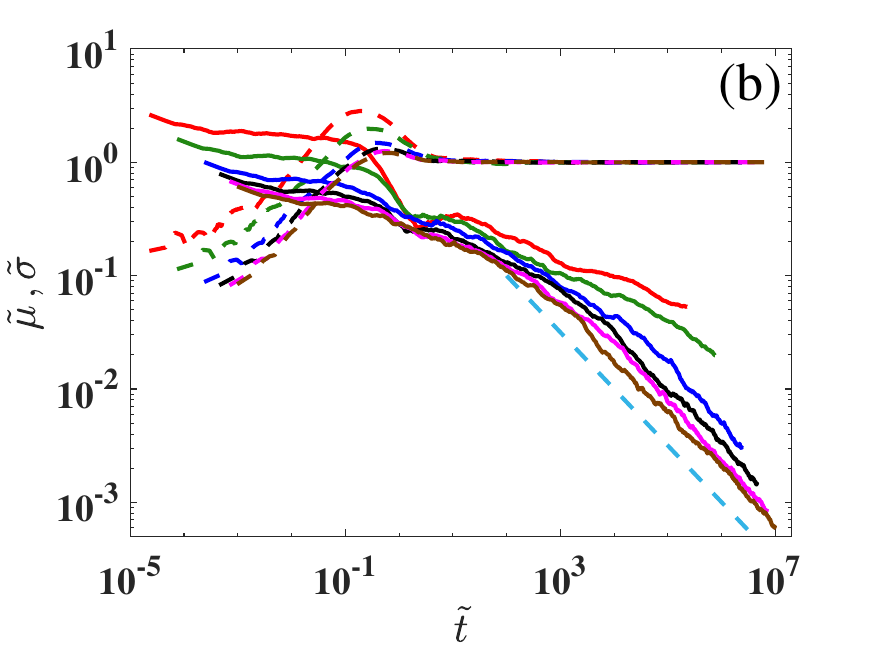}
    \caption{
    The unscaled (a) and scaled (b) average values of the mean and standard deviation of \( \Lambda(t) \) on a log\(_{10}\) scale for the LRN case with \( \theta = 0.33\pi \) for different values of \( g \) (see main text for more information). The colors  red, green, blue, black, magenta, and brown, corresponding to \(g = 0.001, 0.002, 0.004, 0.006, 0.008\), respectively. Averaging is performed over 100 trajectories at each time step. In (b), the cyan dashed line indicates the temporal decay \( \sigma \sim t^{-1/2} \). The time evolution is plotted on a log\(_{10}\) scale, extending up to \( 10^{10} \). The number of unit cells is \( N = 50 \).}
    \label{LRN_N50}
\end{figure}

\begin{figure}[!htbp]
\includegraphics[width=0.45\textwidth]{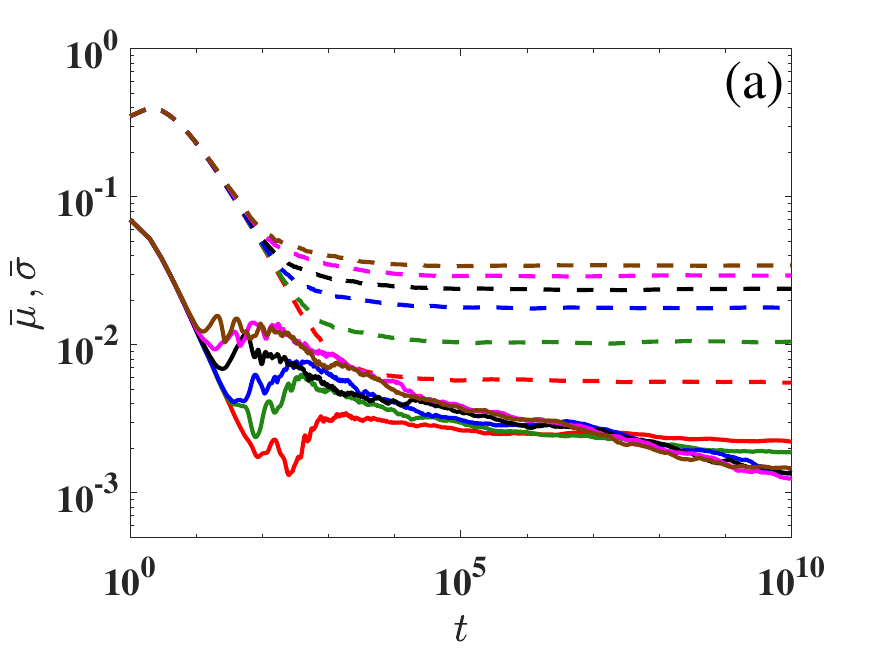}
\includegraphics[width=0.45\textwidth]{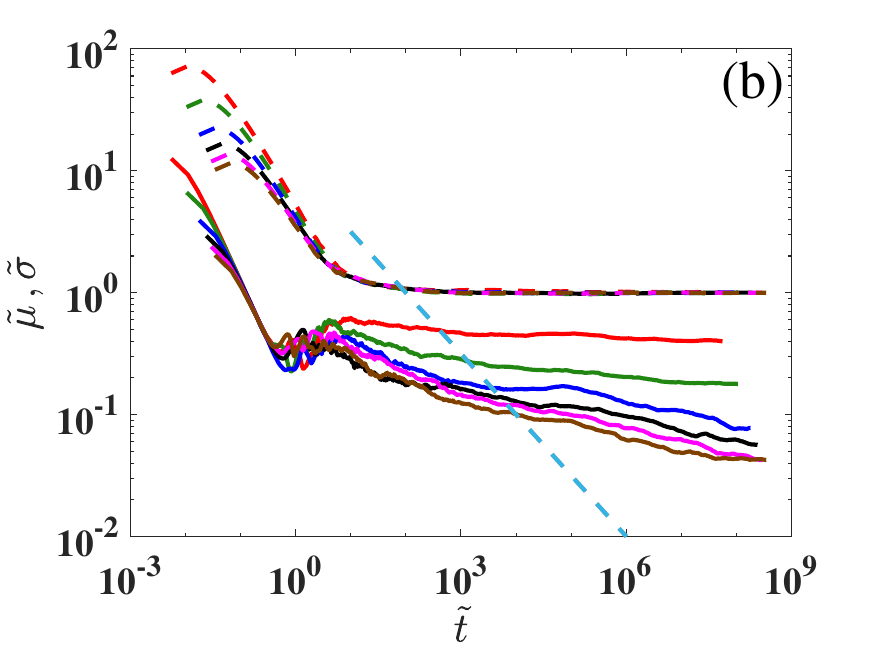}
    \caption{
    Same as FIG.~\ref{LRN_N50} for the SRN case with \( g = 1 \) for different values of \(\theta\). The colors red, green, blue, black, magenta, brown correspond to \(\theta = 0.001, 0.002, 0.004, 0.006, 0.008, 0.01 \), respectively.}
    \label{SRN_N50}
\end{figure}
In order to further quantify our findings, we compute the temporal evolution of the statistical properties of $\Lambda(t)$. After each iteration, we determine the values of $\Lambda(t)$ for each of the 100 trajectories, and compute their mean and standard deviation. We start with the LRN regime illustrated in Fig.~\ref{LRN_N50} (a) for $g=0.01,0.008,0.006,0.004,0.002,0.001$.
The mean deviations show proper saturation, while the standard deviations are decaying with increasing time, indicating asymptotic convergence to proper thermalization and ergodicity. Since the saturation (ergodization) time for each curve is different and roughly inversely proportional to the saturated mean value, we use the saturated values of $z:=\bar\mu (t=10^{10})$
and perform an additional rescaling $\tilde t = zt$, $\tilde \mu(t) = \bar\mu (t) /z$ and $\tilde\sigma (t) = \bar \sigma (t) / z$. The resulting plot in Fig.~\ref{LRN_N50}(b) shows very good merging of the curves for all except the smallest values of $g$ and the standard deviation $\tilde \sigma(t) \sim 1/\sqrt{t}$ as expected and predicted for asymptotic thermalization and ergodicity.

In contrast, the SRN case, shown in Fig.~\ref{SRN_N50}, exhibits a clearly different behavior. In the unscaled data shown in Fig.~\ref{SRN_N50}(a), smaller values of \( \theta \) (e.g., \( \theta = 0.001, 0.002 \)) display a nearly constant \( \bar{\sigma} \) which is almost not decaying with time after \( \bar{\mu} \) stabilizes into a horizontal line. This behavior indicates the presence of a prethermalized state, where fluctuations around the mean persist over long periods, delaying full thermalization. For larger \( \theta \) values (e.g., \( \theta = 0.01 \)), \( \bar{\sigma} \) begins to decay after \( \bar{\mu} \) stabilizes, suggesting faster convergence to thermalization as the coupling strength increases.
Indeed,
the rescaled data in Fig.~\ref{SRN_N50}(b) show an increasingly weaker decay of the rescaled standard deviation when approaching the integrable limit. The decay is much slower than the expected $1/\sqrt{t}$ law shown by the dashed line.  

These results highlight a stark contrast between the LRN and SRN cases. The LRN regime shows consistent thermalization dynamics across all \( g \) values, with \( \bar{\sigma} \) steadily decaying and no evidence of prethermalization. Conversely, the SRN system strongly depends on coupling strength, with smaller \( \theta \) values leading to prolonged prethermalization and delayed thermalization. 

\begin{figure}[!htbp]
\includegraphics[width=0.45\textwidth]{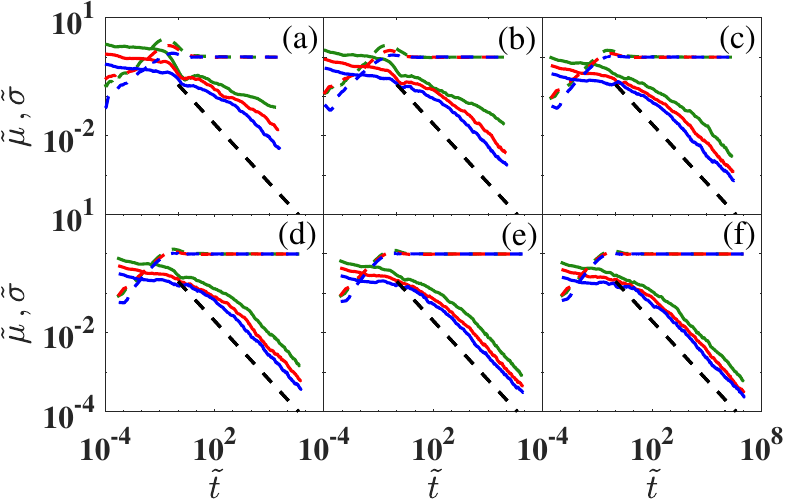}
    \caption{
    Average scaled mean (\( \tilde{\mu} \), dashed lines) and standard deviation (\( \tilde{\sigma} \), solid lines) of \( \Lambda \), calculated from 100 trajectories at each time step, for the LRN model with \( \theta = 0.33 \pi \). Panels (a)–(f) correspond to \( g = 0.001, 0.002, 0.004, 0.006, 0.008, 0.01 \), respectively. The black dashed line shows the expected temporal decay, \( \sigma \sim t^{-1/2} \). Green, red, and blue curves indicate system sizes \( N = 50, 100, 200 \), respectively. All data are plotted on a log\(_{10}\) scale.}
    \label{dif_size_LRN}
\end{figure}

\begin{figure}[!htbp]
\includegraphics[width=0.45\textwidth]{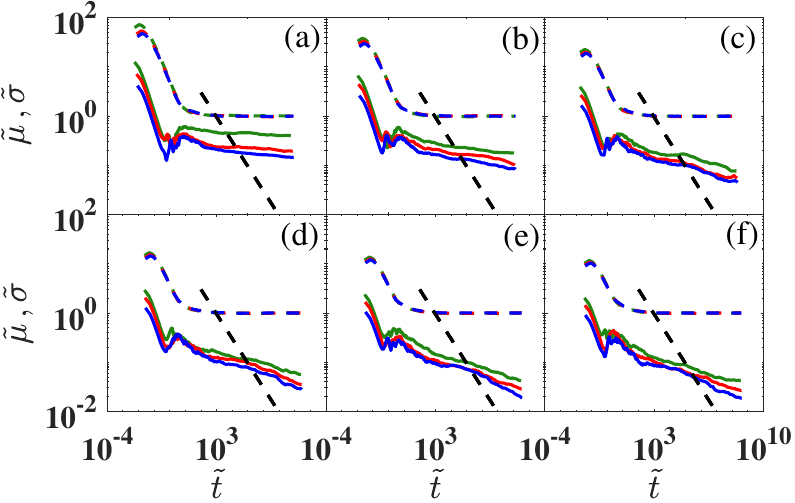}
    \caption{
    Same as Fig.~\ref{dif_size_LRN} for the SRN model with \( g = 1 \). Panels (a)–(f) correspond to \( \theta = 0.001, 0.002, 0.004, 0.006, 0.008, 0.01 \), respectively.}
    \label{dif_size_SRN}
\end{figure}
To investigate the role of system size in the thermalization dynamics, we analyzed the time evolution of \( \Lambda(t) \) for the LRN and SRN regimes and different system sizes, as shown in Fig.~\ref{dif_size_LRN}. and Fig.~\ref{dif_size_SRN}. The analysis includes various system sizes (\( N = 50, 100, 200 \)) for different values of \( g \) (LRN) and \( \theta \) (SRN). The black dashed line represents the theoretical temporal decay \( \sigma \sim t^{-1/2} \), a hallmark of complete thermalization.

For the LRN regime, shown in Fig.~\ref{dif_size_LRN}, the results indicate a deviation from fast thermalization and the slowing down of the $\tilde \sigma$ decay as compared  to the predicted $1/\sqrt{t}$ law, for the smallest size $N=50$ and close proximity to its integrable limit, albeit the effect appears to be rather weak. 
Likewise the SRN regime shows persistent signatures of prethermalization and anomalously slow decay of $\tilde \sigma$, yet no speed-up is either observed for larger system sizes. 

\section{DISCUSSION}

We have investigated the thermalization dynamics of weakly nonintegrable unitary circuit maps, focusing on the statistical properties of Lyapunov observables and the time evolution of the largest Lyapunov exponent. Our analysis reveals distinct thermalization behavior in the SRN and LRN regimes as the systems approach integrable limits.

In the LRN regime, the mean and standard deviation of the Lyapunov observables rapidly converge to zero as the system approaches integrability. This behavior reflects strong ergodicity, and stable thermalization. Meanwhile, the largest Lyapunov exponent quickly stabilizes to a trajectory-independent value, exhibiting typical ergodic thermalization behavior.

In contrast, the SRN regime exhibits pronounced prethermalization phenomena. The statistical properties of the Lyapunov observables demonstrate anomalous fluctuations, with their mean and standard deviation remaining stable over long time scales, forming prethermalization plateaus. Furthermore, the time evolution of the largest Lyapunov exponent shows trajectory-dependent behavior, delaying the system convergence to complete thermalization. These findings highlight the complex dynamics of the SRN regime near integrability and emphasize the significant role of network connectivity in shaping thermalization processes. 

Our results demonstrate that the statistical properties of Lyapunov observables and the largest Lyapunov exponent play a crucial role in characterizing prethermalization and thermalization dynamics in many-body systems. The comparison between the LRN and SRN regimes underscores the profound influence of network topology on the slowing down of thermalization in nonintegrable systems. 

\section*{Acknowledgement}
This research was supported by the Institute for Basic Science
through Project Code (No. IBS-R024-D1). X.Z. acknowledges the financial support from the NSF of China (Grant No. 12247101), the 111 Project (Grant No. B20063), and the China Scholarship Council (Grant No. CSC-202306180087).

\appendix 

\section{On the derivation of short and long range network regimes} 
\label{Resonance}

In the short range network, the integrable limit is reached for $\theta = 0$. The system turns integrable and the equations of motion preserve the local norm (action) $|\psi_n|^2, |\psi_{n+1}|^2$.
For small values of the parameter $\theta$, \(\sin\theta \approx \theta\) and \(\cos\theta \approx 1\), and neglecting higher-order terms like \(\theta^2\), the equations can be approximated up to first order in $\theta$ as

\onecolumngrid 
\vspace{0.5cm}
\noindent\rule{0.5\textwidth}{0.4pt}
\hspace{-7pt}
\raisebox{0cm}[0pt][0pt]{\rule{0.4pt}{0.4cm}}
\begin{equation}
\begin{aligned}
\psi_{n}(t+1) &= 
 e^{i g |\varphi_{n}(t)|^2 }\varphi_{n}(t),\quad  \varphi_n(t) \approx \psi_n(t) - \theta \left[\psi_{n-1}(t) - \psi_{n+1}(t)\right],\\
\psi_{n}(t+1) & \approx 
 e^{i g |\psi_{n}(t)|^2 }\left[ \psi_n(t) -\theta \left\{\left[
 \psi_{n-1}(t) - \psi_{n+1}(t) \right] 
 +ig\psi_n(t) \left( \psi^\ast_n(t)\left[\psi_{n-1}(t) - \psi_{n+1}(t)\right] + \rm{cc}  \right)
 \right\}\right] \;.
\label{A1}
\end{aligned}
\end{equation}
\noindent\hspace*{\columnwidth}\hspace*{-0.5\textwidth}\rule{0.5\textwidth}{0.4pt}
\hspace*{0.5\textwidth}\raisebox{0cm}[0pt][0pt]{\rule{0.4pt}{0.4cm}}
\vspace{0.5cm}
\twocolumngrid
The equations of motion connect the actions through nearest-neighbor interactions and can be categorized as a short-range network.

For the long range network case we switch to the notations from Ref.\cite{malishava_lyapunov_2022,malishava_thermalization_2022} for clarity, that is, replace $\left(\psi_n(t),\psi_{n+1}(t)\right)$ with odd $n$ by $\left( \psi_n^A(t), \psi_n^B(t) \right)$.
For the linear case $g = 0$, we use the standard ansatz $\left( \psi_n^A(t), \psi_n^B(t) \right)^T = e^{-i(\omega_k t - kn)} \left( \psi_k^{A}, \psi_k^{B} \right)^T$, where $\psi_n^A(t)$ and $\psi_n^B(t)$ represent the odd and even parts of $\psi_n(t)$, respectively. The eigenfrequencies $\omega_k$ obey the dispersion relation,
\begin{equation}
\omega_k = \pm \arccos \left( \cos^2 \theta + \sin^2 \theta \cos k \right), 
\label{A2}
\end{equation}
with two dispersive bands $\omega_k^\alpha$ $(\alpha = 1, 2)$ and corresponding normal modes $\boldsymbol{\Psi}_k^\alpha = \sum_n e^{ikn} \psi_k^{\alpha, p} (p = A, B)$, which form a complete set. Generally, a state vector $\boldsymbol{\Psi}(t)$ may be decomposed in terms of normal modes of the linear system,
\begin{equation}
\boldsymbol{\Psi}(t) = \sum_k c_k^\alpha(t) \boldsymbol{\Psi}_k^\alpha. 
\label{A3}
\end{equation}

In the linear case $g = 0$, the evolution of the coefficients $c_k^\alpha(t)$ only involves a phase rotation $e^{i\omega_k^\alpha t}$; the absolute values $|c_k^\alpha|$ are conserved in time, since they are the actions of the integrable limit. Introducing a small nonzero value of $g \neq 0$ results in a coupling between all these actions. Approximating Eq.(\ref{eq1.4}) for small values of $g$, we obtain:

\onecolumngrid 
\vspace{0.5cm}
\noindent\rule{0.5\textwidth}{0.4pt}
\hspace{-7pt}
\raisebox{0cm}[0pt][0pt]{\rule{0.4pt}{0.4cm}}
\begin{equation}
\begin{aligned}
c_k^\alpha(t+1) = e^{i\omega_k} c_k^\alpha(t) + \frac{ig}{N} \sum_{\substack{\alpha_1, \alpha_2, \alpha_3 \\ k_1, k_2, k_3}}
e^{i(\omega_{k_1}^{\alpha_1} + \omega_{k_2}^{\alpha_2} - \omega_{k_3}^{\alpha_3})} I_{k, k_1, k_2, k_3}^{\alpha, \alpha_1, \alpha_2, \alpha_3} c_{k_1}^{\alpha_1}(t) c_{k_2}^{\alpha_2}(t) \left( c_{k_3}^{\alpha_3}(t) \right)^*, 
\label{A4}
\end{aligned}
\end{equation}
\begin{equation}
I_{k, k_1, k_2, k_3}^{\alpha, \alpha_1, \alpha_2, \alpha_3} = \delta_{k_1 + k_2 - k_3 - k, 0} \sum_p \psi_{k_1}^{\alpha_1, p} \psi_{k_2}^{\alpha_2, p} \left( \psi_{k_3}^{\alpha_3, p} \right)^* \left( \psi_k^{\alpha, p} \right)^*. 
\label{A5}
\end{equation}
\noindent\hspace*{\columnwidth}\hspace*{-0.5\textwidth}\rule{0.5\textwidth}{0.4pt}
\hspace*{0.5\textwidth}\raisebox{0cm}[0pt][0pt]{\rule{0.4pt}{0.4cm}}
\vspace{0.5cm}
\twocolumngrid
All $c_k^\alpha$ are coupled due to the second term in Eq.(\ref{A4}). For each action $c_k^\alpha$, the number of terms in the sum is proportional to $N^2$ due to the constraints enforced by the overlap integrals in Eq.(\ref{A5}), resulting in a long-range network.

\section{Further insights into the Lyapunov observable statistics}\label{LYAPOBS}

To provide further details on the behavior of the Lyapunov observable $r(t)$, we present additional numerical results exploring both the SRN and LRN cases with extended parameter ranges closer to the integrable limits, and smaller system sizes to cope with the additional CPU time efforts.
\begin{figure}[!htbp]
\includegraphics[width=0.45\textwidth]{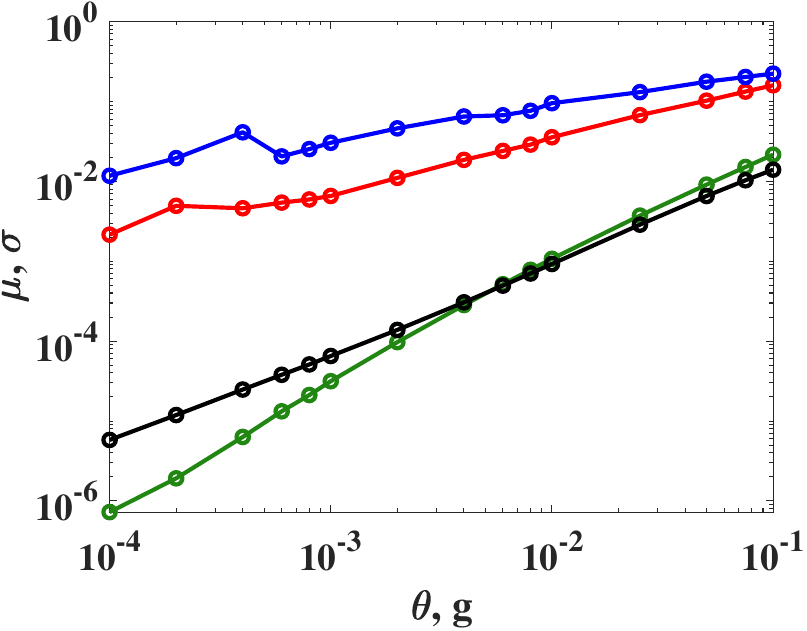}
    \caption{The average value (\( \mu \)) and standard deviation (\( \sigma \)) of the Lyapunov observable for the SRN and LRN cases, with both axes in \(\log_{10}\) scale. The red solid line and blue solid line represent the mean (\( \mu \)) and standard deviation (\( \sigma \)) of SRN with \( g = 1 \) and varying \( \theta \) (\( 10^{-4} \) to \( 10^{-1} \)). The green solid line and black solid line represent the mean (\( \mu \)) and standard deviation (\( \sigma \)) of LRN with \( \theta = 0.33\pi \) and varying \( g \) (\( 10^{-4} \) to \( 10^{-1} \)). The system size is \( N = 50 \), and the evolution time is \( 10^9 \). Similar to Fig.~\ref{muLO}, the averages and standard deviations are calculated using \( 10^9 \) points.}
    \label{mustN50}
\end{figure}
\begin{figure}[!htbp]
\includegraphics[width=0.45\textwidth]{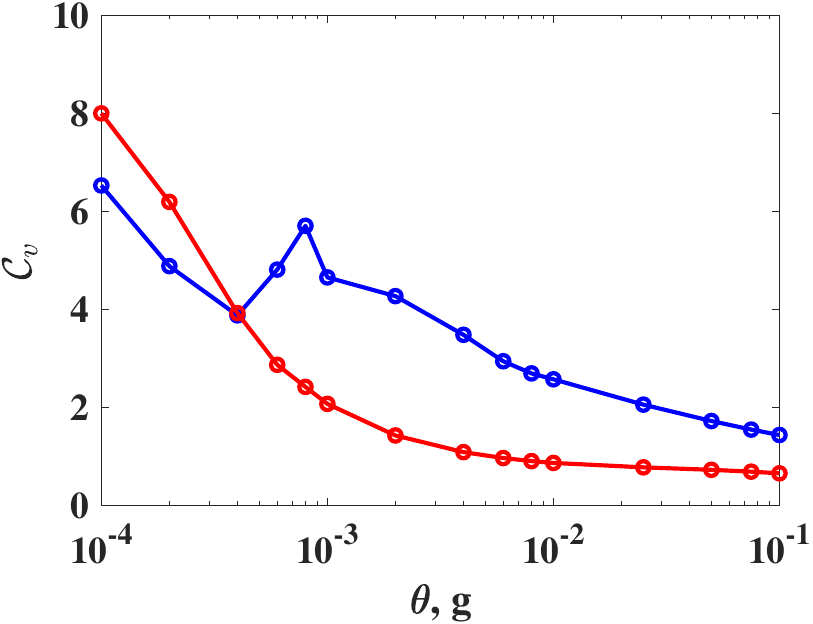}
    \caption{The coefficient of variation (\( \mathcal{C}_v \)) for the Lyapunov observable in the SRN and LRN cases, where \( \mathcal{C}_v  =  \sigma / \mu \) is the ratio of the standard deviation (\( \sigma \)) to the mean value (\( \mu \)). The blue solid line represents the SRN case with \( g = 1 \) and varying \( \theta \) (\( 10^{-4} \) to \( 10^{-1} \)), while the red solid line corresponds to the LRN case with \( \theta = 0.33\pi \) and varying \( g \) (\( 10^{-4} \) to \( 10^{-1} \)). The system size is \( N = 50 \), and the evolution time is \( 10^9 \).
    }
    \label{cv}
\end{figure}

Figure~\ref{mustN50} shows the average values (\(\mu\)) and standard deviation (\(\sigma\)) of the Lyapunov observable distributions. In the SRN case (\( g = 1 \)), as \(\theta\) decreases from \(10^{-1}\) to \(10^{-4}\), both the average and standard deviation appear to gradually saturate to nonzero values, with the standard deviation remaining significantly larger than the mean. This saturation highlights the persistence of nonzero fluctuations near the integrable limit. However, further reducing the value of $\theta$ and approaching the corresponding integrable limit, we observe diminishing of both the mean and standard deviation, as expected.
Yet this decay appears to be very slow. Further studies for larger system sizes and closer distance to the integrable limit are needed in future studies.

For the LRN case (\(\theta = 0.33\pi\)), where \(g\) decreases from \(10^{-1}\) to \(10^{-4}\), both the average (\(\mu\)) and standard deviation (\(\sigma\)) rapidly decrease and appear to converge towards zero. This trend contrasts sharply with the SRN case, where fluctuations persist at smaller values of \(\theta\). At the same time, we note a systematic relative increase of the standard deviation over the mean for $10^{-4}  < g<  10^{-3}$. This might be due to finite size effects. Further studies for larger system sizes and closer distance to the integrable limit are needed in future studies.

Figure~\ref{cv} shows the coefficient of variation \( \mathcal{C}_v \) (defined as \( \mathcal{C}_v = \sigma / \mu \)), which measures the relative fluctuation of the standard deviation to the mean value.
For the LRN case (red line), as \( g \) decreases from \( 10^{-1} \) to \( 10^{-4} \), \( \mathcal{C}_v \) appears to increase monotonically, especially for $g < 10^{-3}$. This indicates that while both the mean and standard deviation approach zero, the relative fluctuations become increasingly significant as the system nears the integrable limit. For the SRN case (blue line), \( \mathcal{C}_v \) is much larger to start with, and also increases when \( \theta > 10^{-3} \), showing a stronger trend but similar to the LRN case. For \( \theta < 10^{-3} \), noticeable fluctuations in \( \mathcal{C}_v \) emerge. We can only speculate about their origins, and repeat again that detailed future studies are needed to clear the fog.

\section{Further insights into the time-dependent Lyapunov exponent for both LRN and SRN } 
\label{LLEDIC}
We provide more details on the time evolution of the largest Lyapunov exponent \( \Lambda(t) \) in the LRN and SRN regimes as they approach their integrable limits.
\begin{figure}[!htbp]
\includegraphics[width=0.48\textwidth]{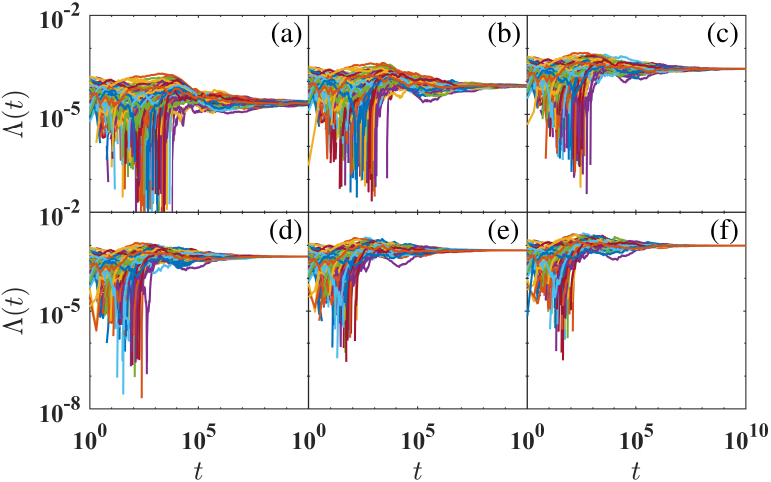}
    \caption{The time evolution of the largest Lyapunov exponent \(\Lambda(t)\) up to \(10^{10}\) 
    for different values of \(g\) for the LRN case (\(\theta = 0.33 \pi\)) within the unitary circuits map. Panels (a) through (f) correspond to \(g = 0.001, 0.002, 0.004, 0.006, 0.008, 0.01\), respectively. Each panel shows 100 trajectories resulting from 100 different initial conditions, displaying the \(\Lambda(t)\) values on a log\(_{10}\) scale. The system consists of \(N = 50\) unit cells.}
    \label{LRN_LLE_N50_A}
\end{figure}

\begin{figure}[!htbp]
\includegraphics[width=0.48\textwidth]{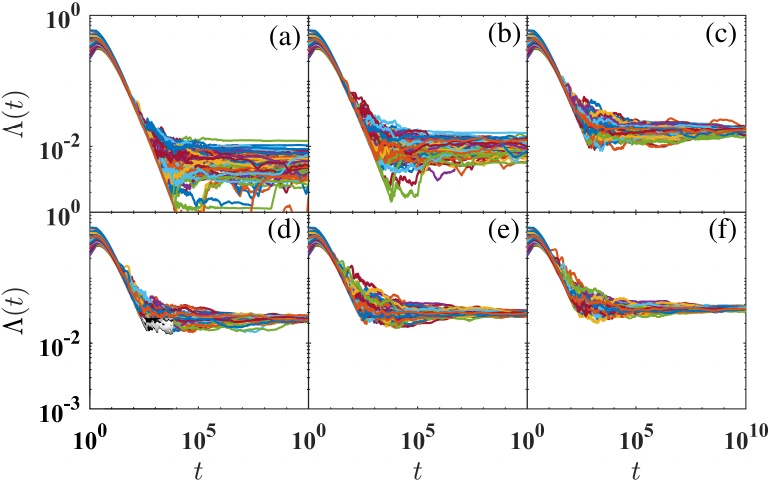}
    \caption{Same as Fig.~\ref{LRN_LLE_N50_A} for different values of \(\theta\) for
    the SRN case with \(g = 1\). Panels (a) through (f) correspond to \(\theta = 0.001, 0.002, 0.004, 0.006, 0.008,0.01\), respectively. The system consists of \(N = 50\) unit cells.}
    \label{SRN_LLE_N50_A}
\end{figure}

Figure~\ref{LRN_LLE_N50_A} shows results for the LRN case with fixed coupling \( \theta = 0.33\pi \) and varying nonlinearity \( g \). Panels (a)-(f) correspond to \( g = 0.001, 0.002, 0.004, 0.006, 0.008, 0.01 \), displaying 100 trajectories with different initial conditions. For large \( g \), \( \Lambda(t) \) quickly converges to a single value, indicating fast thermalization. As \( g \) decreases, convergence slows down, and fluctuations persist longer, but all trajectories eventually stabilize to the same \( \Lambda(t) \). This suggests that prethermalization is absent in LRN systems, and thermalization proceeds robustly even near the integrable limit.

In contrast, Fig.~\ref{SRN_LLE_N50_A} presents results for the SRN case with fixed nonlinearity \( g = 1 \) and varying coupling \( \theta \). Panels (a)-(f) correspond to \( \theta = 0.001, 0.002, 0.004, 0.006, 0.008, 0.01 \). For large \( \theta \), \( \Lambda(t) \) behaves similarly to the LRN case, stabilizing quickly. However, at small \( \theta \), trajectories appear to saturate at distinct values of \( \Lambda(t) \), forming long-lived prethermalization plateaus. These plateaus persist over extended timescales (\( t \sim 10^6 \) to \( 10^{10} \)), indicating a breakdown of delayed thermalization.

Comparing Figs.~\ref{LRN_LLE_N50_A} and~\ref{SRN_LLE_N50_A}, the LRN system shows smooth and rapid thermalization, while the SRN system exhibits prolonged prethermalization with strong dependence on initial conditions. This highlights the role of network structure in thermalization dynamics and establishes prethermalization as a key feature of SRN systems.

\bibliography{bib}

\end{document}